# The Tip-Induced Twisted Bilayer Graphene Superlattice on HOPG: Capillary Attraction Effect

Long Jing Yin[1,§], Wen Xiao Wang[1,§], Ke Ke Feng[1], Rui-Fen Dou[1,*], Jia-Cai Nie[1]

[1] Department of Physics, Beijing Normal University, Beijing, 100875, People's Republic of China

ABSTRACT: We use the tip of the scanning tunneling microscope (STM) to manipulate single weakly bound nanometer-sized sheets on the the highly oriented pyrolytic graphite (HOPG) surface through artifically increasing the tip and sample interaction in humid environment. By this means it is possible to tear apart a graphite sheet againt a step and fold this part onto the HOPG surface and thus generate the gaphene superlattices with hexagonal symmetry. The tip and sample surface interactions, including the van der Waals force, eletrostatic force and capillary attraction force originating from the Laplace pressure due to the formation of a highly curved fluid meniscus connecting the tip and sample, are discussed in details to understand the fromation mechnism of graphen superlattice induced by the STM tip. Especially, the capillary force is the key role in manipulating the graphite surface sheet in the hunmidity condition. Our approach may provides a simple and feasible route to prepare the controllable superlattices and graphene nanoribbons but also replenish and find down the theory of generation of graphene superlattice on HOPG surface by the tip.

Graphene is a strictly two-dimensional sheet of $sp^2$-bonded carbon atoms that can be considered as one atomic layer of graphite. In this perfect two dimensional system, electrons and holes obey to a linear dispersion law and can be described as massless Dirac fermions with a Fermi level coinciding with the Dirac point. This unique and representative band structure results in graphene exhibiting exotic physical properties[1-4]. Since monolayer graphene was mechanically fabricated from bulk graphite in 2004, abundant experimental and theoretical researches of graphene have been launched for exploring this material as a promising component in the next generation of nanodevices[1-7]. Very recently, graphene superlattices become a new center of attention not only because of generation of new Dirac points (superlattice Dirac points, SDPs) in the band structure but also presentation of self-similar recursive energy spectrum induced by the external electronic and magnetic feild[8-12]. It is proved that the new SDPs numbers completely depend on the superlattice potential strength. Hence, graphene superlattices provide a suitable framework to rationally investigate and control the rich band structures and thus physics by changing the moiré superlattice periodicity.

An efficient approach to generate graphene superlattices is to utilize the periodic potential of the substrates, in which monolayer graphene may be placed on top of graphite or hexagonal boron nitride[11-17]. Graphite, a typical stacked layer material, owns the distinct anisotropic property, that is the strong in-plane bonding and weak van der Waals interlayer interactions. Therefore, it is very common to observe moiré patterns on the highly oriented pyrolytic graphite (HOPG) surface by the scanning tunneling microscopy (STM).[14-22] The moiré pattern can also be understood electronically as a superperiodic interference structure: a superlattice, with periods typically in the range of tens of nanometers, akin to the optical interference patterns typically associated with moiré interference. The observed superlattices are largely associated to the native defects produced during the graphite growth or subsequent annealing treatment and so on. On the other hand, these patterns may be from external manipulation by the adhesive tape during cleaning the graphite surface or by the STM tip during scanning the surface [18-24]. Under the external force driving, the topmost layer can be rotated or even folded onto the original graphite surface with a twisted angle relative to the subsurface layer, which result in the superlattices. To date however, there is no very clear explanation on the dynamic process and the interactions between the tip and surface associated with the formation of moiré patterns driven by the STM tip. The earlier literatures reported that the van der Waals force and the electrostatic force between the tip and surface accounted for the tip inducing the topmost layer to be rotated or folded.[20, 24-27] Nevertheless, regarding to the laminated structure stacked by the interlayer weak interaction, Zasadzinski *et al.*, noticed that the tip driving surface layer deformation could more easily occur in the air than in the vacuum[28-30]. They recognized that the strong capillary force arose when the STM tip scanning in the humid air. The capillary forces can increase the tip and surface interaction, which ultimately results in the surface layer deformation such as the changed height and morphology of the scanned areas. Regretfully, this capillary force is seldom considered in the case of the tip folding surface layer to generate superlattices even though not a few of experiments have reported this phenomenon before.

In our work, we carry out a series of studies on the clean HOPG surface subject to the STM tip scanning under ambient conditions at room temperature. We find that the topmost carbon layer can be readily driven to rotate from one twisted angle to another by the STM tip. More unexpectedly, during repeatedly scanning the small-size domains, we *in situ* observe that the surface domain against of the step is tore by the tip and subsequently folded onto the underlying layer to form the moiré patterns. By consideration of the tip and surface interactions including van der Waals force, electrostatic force and the capillary force resulting from the air humidity, the dynamics of the tip tearing and folding carbon layer is rationally interpreted. The driving force originates from the tip-surface interactions, especially the capillary force, that overcome the carbon-carbon bond broken energy, interlayer binding energy and the bending energy. Our experiment not only bring out a simple and cheap route to prepare the controllable superlattices and graphene nanoribbons but also replenish and find down the theory of the tip induced graphene superlattice on HOPG surface.

**METHODS**

HOPG substrates we used in our experiments were of ZYA grade (NT-MDT) and were freshly cleaved prior to experiments using the adhesive tape [22]. MultiMode VIII scanning probe microscope system from Veeco is used in our STM measurement. STM topographic imaging as well as the manipulation experiments with the STM tip is performed in constant current mode under ambient environments at room temperature. The humidity of experimental operation environment can be monitored by the digital thermo hygrometer. The tips used here are PtIr probes which are mechanically formed and purchased directly from Veeco. Typical imaging and manipulating parameters were in the range of 5-50mV for tip bias voltage and 0.7-1.5 nA for tunneling current.

**RESULTS AND DISCUSSION**

Figure 1(a) to (c) show three typical STM images obtained during repeatedly measuring the same area on HOPG surface. The periodical bright dots indicate moiré pattern superlattices in three figures. However, we can obviously find that the periodicity of superlattice structure in three images is not identical during consecutively scanning the same area. It is known that the periodicity ($D$) of the hexagonal superlattice is related to the twisted angle ($\theta$) between the two layers of the hexagonal lattice. In the light of the relationship of the periodicity with the twisted angle by $D = a/[2\sin(\theta/2)]$ ($a = 0.246$ nm is the lattice constant of graphene hexagonal lattice), the periodicities of the superlattices from Figure 1(a) to (c) are 27 nm, 30 nm and 36 nm, while the corresponding twisted angles are 0.52º, 0.47º and 0.38º, respectively. This reveals that structural transition occurs during repeatedly scanning the same area for half an hour. According to the variation in the twisted angles between the top and underlying layer, the schematic models of the initial and final superlattice structures with the respective twisted angles of 0.52 º and 0.38 º are shown in Figure 1(e) and (f). These figures display two different moiré patterns, in good agreement with the Figure 1(a) and (c).

Here the structural transition can be attributed to the topmost layer being driven by the STM tip to rotate related to the underlying layer, which generates the superlattices with the different twisted angles. Actually, it has reported that the extremely low sliding friction between graphene layers makes graphite as an extremely good lubricant.[31-33] Thus the topmost layer is easily induced by the external forces such as the tip and surface interactions to laterally displace or twist with respect to the underlying layer. The interactions between the STM tip and sample surface normally comprised of the van der Waals interaction and the electrostatic force, which lead to the generation of the structural transition. However, since the capillary force between the tip and sample surface is inevitable during scanning the laminated structure sample in air, what is it contribution to the superlattice structural transition from the capillary force? Here we should clarify the transition mechanism more deeply and completely.

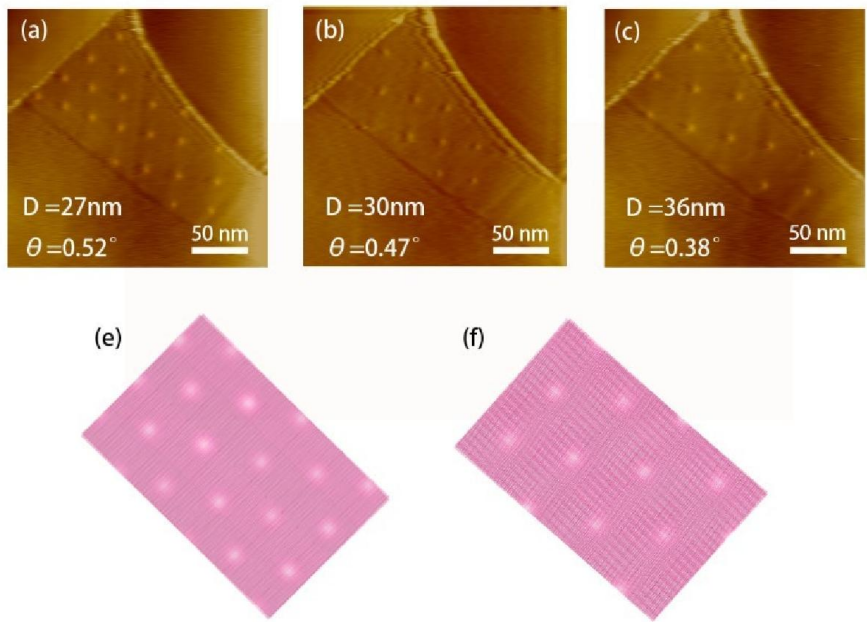

**Figure 1.** (a)-(c) Three typical STM images of the clean HOPG surface from the same scanned region during repeatedly scanning for half a hour. (a) In the first stage, the period of moiré pattern $D_a$ = 27 nm, corresponding to the twisted angle $\theta$ = 0.52°. (b) After scanning 15mins later, the period and twisted angle of the moiré pattern are changed into $D_b$ = 30 nm and $\theta$ = 0.47°, respectively. (c) In the final stage followed by image b, the period and twisted angle changed into $D_c$ = 36 nm and $\theta$ = 0.38°. Scanning parameters for (a), (b) and (c): Vs = 25 mV, and I = 1 nA;

Scanning size: 230×230 nm$^2$. (e) and (f) Schematic models of two moiré patterns with the variation in the periods and twisted angles.

More unexpectedly, when keeping on scanning the fixed domain adjoin to a step of graphite in the constant current mode for longer interval, we *in situ* observe that a nanometer scale topmost surface is tore apart by the tip and then folded back onto the graphite substrate with a small rotation angle. The tore area is labeled by the blue dash closed loop in Figure 2(a) to (c). Compared Figure 2(b) with 2(c), except that carbon atoms located in the edge is split and eventually disappeared from the torn surface region during the manipulation process, as labeled by characters "A" and "B", the basic characteristics remain as before during the consecutively scanning (See more structural information in Supporting Information). As a result, a pristine moiré pattern in the tore area disappears while a new moiré pattern forms (labeled by the letter "F" in Figure 2(b)). The periodicity of the pristine moiré pattern ( the area "D") is about 7 nm with the corresponding twisted angle of 2º, while the superlattice of the new superlattice (the area "F") is changed to 5 nm with the twisted angle of 2.8º. For more intuitively understanding the manipulation process, the schematic models are shown in Figure 2(d).

Observations of moiré superlattices on graphite induced by the STM tip have been reported in the literatures.[34-36] In the earlier experiments, Bernhardt and Gan *et. al*. have provided a similar approach of preparing a superlattice on graphite artificially. However, there is short of deep interpretation on what is the driving force to generate superlattices. Here we will discuss all possible interactions between the tip and surface to understand the transition mechanism of the tip manipulating the surface structures to be tore and even folded. As we have mentioned above that the tip driving surface layer deformation could more easily occur in the air than in the vacuum [28-30], the strong capillary force cannot be neglected in the case of STM experiments performed under the humidity conditions. Moreover, as is a general case for organic films or even layered crystals with weak interlayer interaction, the tip and surface interactions can be described as a schematic model (Figure 3).

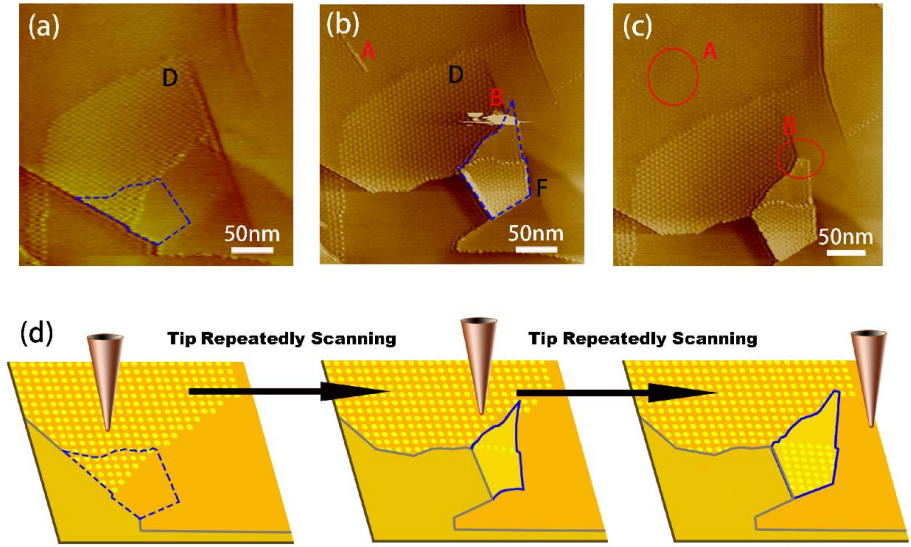

**Figure 2.** (a)-(c) *In situ* STM images of the same scanning region during the scanning interval exceeding half an hour. Scanning size is all $350 \times 350$ nm$^2$. (a) First scanning process near a graphene step edge; (b) After 10mins scanning later; (c) After 20mins scanning later; (d) Explanation of tip-induced method by using the STM tip to form a moiré patterns superlattice. Scanning parameters: Vs = 25 mV, I = 1 nA and Scan Rate = 2.96 Hz.

According to this model, there are at least three significant interactions to be considered: (1) van der Waals attraction; (2) electrostatic interaction due to the potential difference between the tip and sample; (3) capillary attraction due to the Laplace pressure generated by the formation of a highly curved fluid meniscus connecting the tip and sample[28-30].

For a spherical tip of radius R and a tip-sample surface distance D, the van der Waals force[25] is

$$F_{vdW} = \frac{AR}{6D^2}.$$

Where A is the Hamaker constant between the tip and the surface, normally equals to $1 \times 10^{-19} J$.[37] Hence, $F_{vdW}$ is in the range of $0.1 \sim 1 nN$, for *R* is about 15 ~ 30 nm, and *D* is about 1nm. The electrostatic force between a sphere tip at a potential, $V_0$, and a grounded sample is given by below:[38]

$$F_E = \pi \varepsilon V_0^2 R / D.$$

where $\varepsilon$ is the constant of electronic and $V_0$ is the bias between the sample and the tip. When the relationship of the tip radius (R) with the tip-sample distance (D) meets the condition: $R \gg D$, and $V_0$ is about 0.1~3V, $F_E$ is in the range of $0.01 \sim 10 nN$.

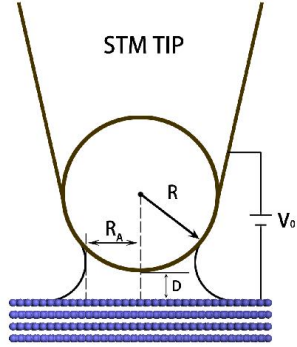

**Figure 3.** Schematic model of a fluid meniscus connecting the STM tip with the sample surface.

In the present case, we scan the HOPG sample in the ambient and humid conditions. Thus a highly curved fluid meniscus easily forms to connect the tip and sample. This implies that the Laplace pressure can give rise to due to this curved fluid meniscus. The capillary force[39-41] due to the meniscus between the sample and tip is

$$F_{cap} = \pi R_A^2 \frac{R_g T \ln(RH)}{V_m}.$$

Where $V_m$ is the molar volume of the liquid solution (correspond to water in this experiment), $R_g$ is the ideal gas constant ($R_g$ =8.314472 J·mol$^{-1}$·K$^{-1}$), $RH$ is relative humidity, and $T$ is the experimental temperature ($T$ = 298K) and $R_A$ is the inner radius of the capillary neck between tip and sample surface. Given that $R_A$ is normally less than the radius of the tip,[42] and the relative humidity is over than 10%, we provide a summary in Table 1 for the above three interactions between the tip and sample surface.

From the above table, we can find that the order of the capillary force is much larger than that of other two. After considering two kinds of relative humidity, the capillary force will be changed dramatically, as shown in Table 2. This further provides direct evidence that the humidity plays a key role on increasing the tip and surface interaction force. Nevertheless, it is worth noting that the capillary force could decrease with increasing relative humidity over than 30%.[43] In our experiment, the relative humidity is controlled between 20% and 30%. Under these circumstances, the total force between the tip and sample surface is up to 34.6 nN under the relative humidity of 20%, which is close to 32 KeV. Moreover, we find that the total energy is mainly from the capillary force. Then we want to identify whether this total energy between the tip and sample surface can provide a driving force to tear and eventually fold the topmost layer?

| The different forces between the tip and sample surface | Functions | The orders of the different forces |
|---|---|---|
| van der Waals force | $F_{vdW} = \dfrac{AR}{6D^2}$ | $0.1 \sim 1 nN$ |
| Electrostatic force | $F_E = \pi \varepsilon V_0^2 R / D$ while $R \gg D$ | $0.01 \sim 10 nN$ |
| Capillary force | $F_{cap} = A_m \Delta P_L = \pi R_A^2 \dfrac{R_g T \ln(RH)}{V_m}$ | $10 \sim 100 nN$ |

**Table 1** A summary of three interaction forces between the tip and sample surface.

| Interaction | $R_H = 0.10$ | $R_H = 0.20$ |
|---|---|---|
| $F_{vdW}$ | 0.3 | 0.3 |
| $F_{ele}$ | 0.01 | 0.01 |
| $F_{cap}$ | 25.7 | 34.3 |

**Table 2** A comparison of calculated three forces in two kinds of relative humidity environment (in nN). ( $R = 20$ nm, $D = 1$ nm, $V = 30$ mV)

We recall our experimental result by Figure 2(b). This image shows that this graphene sheet against a step is tore apart from the intact surface layer and folded onto the remnant area during repeatedly scanning for longer interval. A new graphene superlattice pattern appears on the scanned area as labeled by the letter "F" in Figure 2(a) to (c). In the whole manipulation process, the energy dissipation is associated to the energy of the carbon-carbon (C-C) bond broken, graphene interlayer binding energy from two neighboring graphene layers and curvature energy from the torn layer folded onto the topmost surface. As for breaking the C-C bonds, there are two cases, breaking bonds along the zigzag direction and along the armchair direction, as displayed in Figure 4(a) and (b). According to the number of the C-C bond broken in the tore graphene area, the C-C bond broken energy can be evaluated by the function:

$$E_{bond} = N'E^0_{bond}$$

$N'$ is the broken C-C bond number and $E^0_{bond}$ is the C-C bond energy of 4.9 eV[44]. The energy of the C-C bond broken is approximate 3.7 KeV and 3.0 KeV for the armchair and zigzag edges, respectively. The curvature energy can be evaluated based on the following expression [45-46]:

$$E_C = (\frac{1}{2})K(\frac{1}{R'})^2 A$$

where $A = \pi R'L$, $K = 1.4$ eV[45], $R'$ is the curving radius and $L$ is the length of the curving graphene sheet. We obtain these constants according to the model of Figure 4(d). The calculated curvature energy is approximate 154 eV. Besides two kinds of energy, the interlayer binding energy can be calculated by

$$E_{bind} = NE^0_{bind}$$

In the present case, $N$ is the C atom number and $E^0_{bind}$ is the graphite interlayer binding energy. $E^0_{bind}$ is theoretically and experimentally defined as 45.5 meV /atom.[19,47-48] Even though the graphene interlayer binding energy is controlled by the weak van der Waals interaction, the counted binding energy related to the manipulated area is about 10 KeV. The above three kinds of energy are summarized in the histogram, shown in Figure 4(e). It reads that the curvature energy is much less than two others. The total energy barrier is about 14 KeV, which is required to be surmounted to break the C-C bond, bend the graphene layer and fold the layer onto the topmost graphene layer surface to produce the superlattices. Accordingly, the interactions between the tip and sample surface should be large enough to overcome this energy barrier to manipulate the graphene flakes. In our case, the total energy between the tip and sample surface is about 32 KeV, which is much larger than the energy barrier of 14 KeV. As a result, the tip can provide a driving force to tear the graphene layer and fold the layer onto the remnant surface. Moreover, among the three kinds of the tip-surface force, the capillary force originating from the curved fluid meniscus connecting the tip and surface is the main role in manipulating the topmost surface when scanning the sample in ambient condition. The capillary force is blindingly obvious in our experiment due to the HOPG sample measured in ambient and humid environment. In other words, our explanation on the manipulation of the topmost layer in the layer-like surface is universal in the event of experiments being performed in ambient and even humidity environment because the curved fluid meniscus connected the tip and surface cannot be unacknowledged.

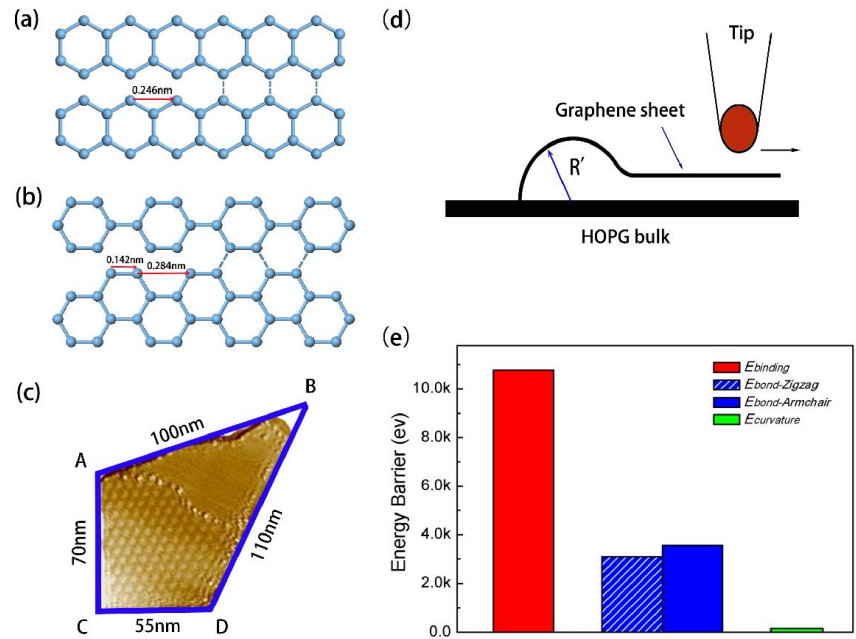

**Figure 4.** (a) Bond broken along the zigzag direction. The distance is 0.246 nm between two C-C bonds. (b) Bond broken along the armchair direction. The distance is 0.426 nm between two C-C bonding. (c) Typical STM image representing a torn and folded graphene sheet. Lines of AB and CD were two cleaving edges. (d) Schematic model of the graphene sheet being folded. R′ is the curving radius. (e) Histogram of three energy barriers, i.e. bending energy (green pillars), graphene interlayer coupling energy (red pillars) and the energy of C-C bonds being broken (blue pillars).

In conclusion, We use the STM tip to manipulate weakly interlayer bound nanometer-sized sheets on HOPG surface through aritfially induce a strong tip-sample interaction, a capillary attaction force by performing experiments in the humid condition. Using this means it is possible to tear apart a graphite sheet against a step and fold onto a HOPG surface region and thus create superstructures with hexagonal symmetry. Especially, the capillary force generated by the Laplace pressure duo to formation of a highly curved fluid meniscus connecting the tip and sample is a driving force to manipulate the graphite surface sheet to be tore and folded in the hunmidity conditions. Our approach may provides a simple and feasible method to produce the superlattice structures on HOPG surface. In addition, given the weak electronic coupling between the layers of graphene in graphite, studies of superlattices can provide a suitable framework to rationally investigate and control the rich band structures and thus physics by changing the moiré superlattice periodicity.

**Acknowledgements**

This work was supported by the Ministry of Science and Technology of China (Grants No.2012CB933022) and the National Natural Science Foundation of China (Grant No. 10804010).

**Author contributions**

L.J.Y., W.X.Wang have the same contribution to this work.

**Correspondence Authors**:

rfdou@bnu.edu.cn

# Supporting Information

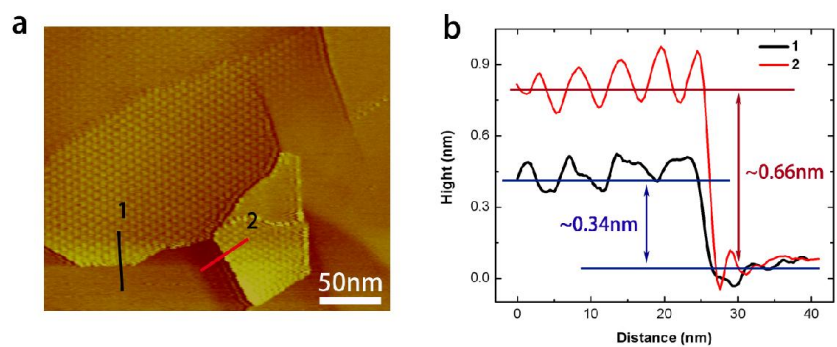

Figure S1 (a) STM image shown in Figure 2. (b) Profiles along the section lines 1 and 2 shown in panel **a**.

Figure S1(a) shows ae STM topological image of the folded graphene sheet. The height of the original graphene step is about 0.34 nm, displayed in Figure S1(b), which corresponding to one graphene interlayer distance. The high difference of the profile line 2 is around 0.66 nm, demonstrating that a graphene domain was folded onto the original sheet induced by the STM tip.